\documentclass{ifacconf}

\usepackage{graphicx, color, amsmath, url, amssymb}   
\usepackage{natbib, booktabs}        

\usepackage{tikz}  
\usepackage{sidecap} 

\newtheorem{ass}{\sc Assumption}
\newcommand{\ader}{\mathcal A_\text{der}}
\renewcommand{\e}{\vskip 1mm\noindent}
\begin{document}
\begin{frontmatter}

\title{On Continuous-time Sparse Identification of Nonlinear Polynomial Systems} 


\author[First]{Mazen Alamir} 

\address[First]{Univ. Grenoble Alpes, CNRS, Grenoble INP, GIPSA-lab, 38000 Grenoble, France  (e-mail: mazen.alamir@grenoble-inp.fr).}

\begin{abstract}                
This paper leverages recent advances in high derivatives reconstruction from noisy-time series and sparse multivariate polynomial identification in order to improve the process of parsimoniously identifying, from a small amount of data, unknown Single-Input/Single-Output nonlinear dynamics of relative degree up to $4$. The methodology is illustrated on the Electronic Throttle Controlled automotive system. 
\end{abstract}


\end{frontmatter}
\section{Introduction}
In this paper we are interested in  continuous-time identification  of unknown Single-Input/Single-Output (SISO) polynomial systems of relative degree $n\in \{1,\dots,4\}$. Namely:
\begin{equation}
y^{(n)}=\mathcal P(y,\dot y, \dots, y^{(n-1)}, u) \label{siso}
\end{equation}
where $\mathcal P$ is a multivariate polynomial of an unknown degree $d$. Moreover since the \textit{safe} excitation of unknown systems remains challenging, we are interested in a methodology that enables an accurate identification using small datasets involving a family of short damped excitation signals that are known to be possible to safely apply to the system, initially at rest.
\e 
While the restriction applies to the notation and the specific illustrative example, the methodology developed hereafter easily extends to more general MIMO polynomial systems of the form: 
\begin{align}
 &\text{For}\ i\in \{1,\dots, n_y\},\nonumber\\
 &y_i^{(n_i)}=\mathcal P_i\Bigl(\bigl\{\{y_j^{(\kappa)}\}_{\kappa=0}^{n_j^{(i)}-1}\bigr\}_{j=1}^{n_y}, \{u_\ell\}_{\ell=1}^{n_u}\Bigr)=:\mathcal P_i\bigl(z_i\bigr)\label{mimo}
\end{align}
using the same tools that are shown hereafter to be efficient to address the identification of \eqref{siso}, namely:
\begin{itemize}
\item[-] \texttt{Tool1}: The recently proposed algorithm\footnote{This refers to the \texttt{python} module \texttt{ml-derivatives} that is available through standard \texttt{python-pip-install}. see \url{https://mazenalamir.github.io/ml_derivatives/} for more details.} for the computation of high derivatives (up to order $4$) of noisy time-series \citep{alamir2025reconstructing},\\
\item[-] \texttt{Tool2}: The scalable \texttt{plars} algorithm which addresses the problem   of identifying multivariate polynomial relationships \citep{MazenBookplars2025}.
\end{itemize}
Indeed, assuming that for a given $n_i$, the l.h.s $y^{(n_i)}_i$ of \eqref{mimo} is available (\texttt{Tool1}), the problem boils down to identify the multivariate polynomial relationships \eqref{mimo} that links the \texttt{label} $y_i^{(n_i)}$ to the \texttt{features}\footnote{The couple \texttt{(label, features)} borrows the well-known Machine Learning vocabulary where one seeks a map $F$ such that \texttt{label}=F(\texttt{features}).} vector $z_i$ (\texttt{Tool2}). 
\e The price associated to the generalization to the MIMO systems lies in the dimension of $z_i$, namely\footnote{Notice that $n_i^{(i)}=n_i-1$ for all $i$ by definition.}: 
\begin{equation*}
\texttt{dim}(z_i) = n_u + \sum_{j\in \{1,\dots,n_y\}}n_j^{(i)}
\end{equation*}
as this leads to a multivariate polynomial with a large number of coefficients. For instance, a \textbf{9-dimensional} nonlinear system with two inputs and three outputs of relative degree\footnote{The relative degree of an output is the number of derivation that are required to see the one of the components of the input explicitly in the expression of the derivative.} $3$ each, might be associated to the following instantiation: 
\begin{equation*}
\Bigl\{n_u=2, n_y=3, n_j^{(i)}=3\Bigr\}\ \longrightarrow \texttt{dim}(z_i)=11
\end{equation*}
This leads to a 78 coefficients for a polynomial of degree $d=2$, $1365$ coefficients for $d=4$ and $12376$ coefficients for $d=6$. 
\e Surprisingly enough, as it is shown in \citep{MazenBookplars2025}, such problems can be efficiently solved using standard tools as the \texttt{lassolars} module of the scikit-learn python library \citep{scikit-learn} in few minutes if not seconds. Much larger problems involving up to half a million variables can be solved using \texttt{plars} \citep{MazenBookplars2025} in few minutes suggesting that this is not a real issue.
\e 
This being said, for the sake of simplicity of notation, we focus on SISO models given by \eqref{siso} in the present paper. 
\e
The reason for which we are interested in continuous-time models is threefold: \begin{enumerate}
    \item It enables to totally decouple the identification of the dynamics from the choice of the sampling period for the control design. \\
    \item More importantly, the degree of the r.h.s of the continuous-time dynamics is lower than that of the discrete-time one-step ahead version as the latter is the result of multiple composition of the former as it is the case in standard Runge-Kutta method for instance. The direct consequence of this fact is that it is an easier task to identify the ODEs than the discrete-time law, provided that the reconstruction of high derivatives is possible despite of the noisy measurement (\texttt{Tool1}).  \\
    \item Identifying the nonlinear system in the derivatives space corresponds to a nonlinear coordinates transformation that might, in some cases, approximately admits a linear representation enabling simpler design of the control law. This is because the implicit coordinates change involves the r.h.s of the ODE's. This features is precisely exhibited in the illustrative example used in the present paper. 
\end{enumerate}
It is precisely because decent reconstruction of high derivatives has always been viewed as \textit{almost-impossible} that CT-identification was systematically considered as a tricky task \citep{rao2006identification, PADUART2010647}. 
\e 
In the next section, a discussion regarding possible alternatives to address the above-defined problem is briefly provided
\subsection{State of the art}
A relevant way to cluster the set of possible approaches is to examine the mathematical structure that is used to represent the nonlinearity:
\e 
- The \textbf{Hammerstein} structure \citep{greblicki2002hammerstein, JUI2021339} has been widely used in the past where a nonlinear transformation of the control is fed to a linear dynamics. This results in a nonlinear structure that is affine in the state and can be nonlinear only in the control input. 
\e 
- The \textbf{Wiener} structure \citep{sadeghi2018online} where nonlinearity applies to the output of a linear dynamics, the nonlinearity is affine in the control and all the state's components except the output itself which is again a very specific structure. 
\e 
- Naturally, combined \textbf{Wiener-Hammerstein} structures have been studied in \citep{paduart2012identification} with polynomial nonlinear part. This leads to the nonlinearity affecting only input and output and not the internal state or, equivalently, the other derivatives of the output as in \eqref{siso}. 
\e 
- Comprehensively, the recent burst of \textbf{Deep NN} enhanced some works that addressed the identification of the two above structures using DNN to express the localized nonlinearity but this does not fundamentally change the  properties of the nonlinearity being exclusively concentrated on the input and the output respectively. 
\e 
- \textbf{Fully-polynomial state space models} have been studied in \citep{PADUART2010647} and some related papers. The algorithm in \citep{PADUART2010647} involves two main steps. In the first a \textit{best} linear model is first identified. This model is then \textit{corrected} by adding virtual inputs that involve a set of multivariate monomials in the state and the control input. The identification of the nonlinear added polynomial expressions is based on the minimization of the norm of the the output prediction error which involves the simulation of the nonlinear system for all candidate set of parameters involved in the iterations. Since neither the linear model nor the successive nonlinear iterates are guaranteed to show stable dynamics, this might lead to a fragile optimization process. 
\e 
This difficulty is shared by all the algorithms that minimize \textbf{simulation-based} cost functions. Another difficulties stems from the \textbf{non parsimonious} approach being used which might induce the well known \textit{over-fitting} problem unless a large dataset is used which contradicts the problem statement as defined above.
\e 
In \citep{diagetti19}, a Machine Learning approach is proposed that involves the \textbf{Particle Bernstein Polynomials} (\textbf{PBP}). The output is viewed as a nonlinear (PBP) function of the components of the input trajectory on some modal basis. The nonlinear function is identified via standard ML regression technique after Principal Component Analysis (\textbf{PCA})-based dimensionality reduction. While the approach is discrete-time oriented it can be adapted to continuous-time setting by considering the label to be the higher derivatives of the measured outputs. Notice however that the modal approach needs extensive dataset.
\e 
Solution involving \textbf{Reinforcement learning} \textbf{(RL)} can be imagined where the nonlinear map in \eqref{siso} would be identified via standard RL episodes. In each episode, a virtual chain of integrators is \textbf{simulated} using the current iterate of the DNN representing the r.h.s of \eqref{siso} and the reward is computed based on the output prediction error. Notice however that being a simulation-based approach, this solution would share the same stability-related issue discussed above leaving aside the non sparse character that would require an important amount of learning data in order to avoid overfitting on the small learning dataset that is mentioned in the problem's statement.
\e
Much more closer to the present work, the use of time derivatives to perform continuous-time \textbf{sparse identification} of nonlinear systems has already been proposed in \citep{brunton2022data} and the related works among which, a python package \citep{de2020pisindy} is made available. However the framework is based on \textbf{first derivatives} of all the state components that are supposed to be fully measured. This assumption \textit{works around} the reconstruction of higher derivatives at the price of a strongly restrictive, if not unrealistic, assumption. 
\e 
This quick overview only sketches how vast this topic is and suggests that there is no free-lunch like solution to the problem that outperforms the others in all circumstances. The proposition made in the present contribution add a slightly different step which builds on recent significant advances in related topics and tools. 
\e 
This paper is organized as follows: First, some recalls and definitions are proposed in Section \ref{sec-recalls}. Section \ref{sec-method} presents the identification methodology that is illustrated in Section \ref{sec-illustration} using the automotive ETC as a supporting example. Finally, Section \ref{sec-conc} concludes the paper and gives hints for further investigation.

\section{Some definitions and recalls}\label{sec-recalls}
First of all, since this contribution is about multivariate polynomials, some related recalls and notation are necessary. 
\e 
A multivariate polynomial in $z\in \mathbb R^{n_z}$ takes the form: 
\begin{equation}
\mathcal P(z)=\sum_{i=1}^{n_m}c_i\phi_i(z)\ \text{where}\ \phi_i(z)=\prod_{j=1}^{n_z}z_j^{p_{ij}} \label{defdePz}
\end{equation}
where $\phi_i$ is referred to as the $i$-th monomial of $\mathcal P$. The integer $n_m$ refers to the number of monomials used in $\mathcal P$. Consequently, a polynomial $\mathcal P$ is totally defined by the pair $(P,c)$:
\begin{equation}
P := \Bigl\{p_{ij}\Bigr\}\in \mathbb N^{n_m\times n_z}\quad,\quad  c\in \mathbb R^{n_m} \label{defdeP}
\end{equation}
representing respectively the matrix of monomial powers and the associated coefficients. 
\e The degree $d_i$ of a monomial $\phi_i$ is defined by $d_i=\sum_{j=1}^{n_z}p_{i j}$. The degree of the polynomial $\mathcal P$ is the maximum degree of its monomials with non vanishing coefficients $c_i$, namely  $d=\max_{i=1}^{n_m}\Bigl\{d_i\ \vert\  c_i\neq 0\}$. Given the dimension $n_z$ of $z$ and the degree $d$ of the polynomial, the number $n_m$ of candidate monomials is given by\footnote{This can be computed using the \texttt{python, math.comb} module.}: 
\begin{equation}
n_m=\begin{pmatrix}
n_z+d\cr d  
\end{pmatrix} \label{expressionofnc}
\end{equation} 
\e 
The framework proposed in this paper is based on the following assumptions: 
\begin{center}
\begin{tikzpicture}
\node[fill=black!5, rounded corners, inner xsep=3mm, inner ysep=4mm](T){
\begin{minipage}{0.43\textwidth}
\begin{ass}
It is assumed that there is a set of input profiles, denoted by $\mathbb U_\text{safe}\in \mathbb R^{N}$ such that when $\mathbf{u}\in \mathbb U_\text{safe}$ is applied to \eqref{siso}, initially at rest, no instability occurs so that the resulting measurement can be \textit{safely} collected for use in the identification. 
\end{ass}
\end{minipage} 
};
\node[above] at(T.north){\small \sc $\exists$ Safe excitation inputs from rest};
\end{tikzpicture}
\end{center} 
Notice that $\mathbb U_\text{safe}$ does not necessarily match the conditions required by standard (linear, dense) identification theory in terms of bandwidth. This is where sparse identification becomes mandatory.
\e The next assumption concerns the availability of high-derivatives estimation algorithm:
\begin{center}
\begin{tikzpicture}
\node[fill=black!5, rounded corners, inner xsep=3mm, inner ysep=4mm](T){
\begin{minipage}{0.43\textwidth}
\begin{ass}
There exists a self-tuning algorithm $\mathcal A_\text{der}$, such that for all $n\in \{0,1,2,3,4\}$ and all time-series $\mathbf{v}\in \mathbb R^N$ representing $\tau$-sampled successive values of a smooth signal, the call:
\begin{equation}
(\hat y^{(n)}, \sigma^{(n)}) \leftarrow \ader (\mathbf{v}, n, \tau)\label{aderformulae}
\end{equation}
provides an estimation $\hat y^{(n)}\in \mathbb R^N$ of the $n$-derivatives of the time-series together with a sequence of confidence indicators $\sigma^{(n)}\in \mathbb R^N$.
\end{ass}
\end{minipage} 
};
\node[above] at(T.north){\small \sc $\exists$ an algorithm for Higher derivatives reconstruction};
\end{tikzpicture}
\end{center} 
As a matter of fact, such an algorithm ($\ader$) has been recently made freely available through the release of the \texttt{ml-derivatives} Python package which implements the framework proposed in \citep{alamir2025reconstructing}. Notice that the commonly encountered tricky to determine trade-of between the level of noise and the filtering order of the reconstruction is automatically handled by the algorithm so that only the input arguments $(\mathbf{v}, n, \tau)$ invoked in \eqref{aderformulae} have to be provided. The limitation to $n\le 4$ is simply linked to the current version of the implementation and will be shortly extended to higher derivation orders. 
\e 
The last assumption concerns the algorithm performing sparse identification of multi-variate polynomials: 
\begin{center}
\begin{tikzpicture}
\node[fill=black!5, rounded corners, inner xsep=3mm, inner ysep=4mm](T){
\begin{minipage}{0.43\textwidth}
\begin{ass}
There exists an algorithm $\mathcal A_\text{id}$ such that, given a features matrix $Z\in \mathbb R^{n_s\times n_z}$ involving $n_s$ samples and a corresponding vector of targeted label $\ell$, the call:
\begin{equation}
\mathcal P:=(P, c) \leftarrow \mathcal A_\text{id} (Z, \ell, \varepsilon)\label{aidformulae}
\end{equation}
provides the multivariate polynomial  $\mathcal P:=(P,c)$ which is the sparse solution to the least squares problem: 
\begin{equation}
\min_{\mathcal P} \sum_{j=1}^{n_s} \left\|\ell_j-P(z_j)\right\|^2  \label{leastdsq}
\end{equation}
where $\ell_j$ and $z_j$ stand for the $j$-th row of $\ell$ and $Z$ respectively. The optional parameters $\varepsilon$ might be used as a threshold defining the relevance of adding more monomials to the current solution. 
\end{ass}
\end{minipage} 
};
\node[above] at(T.north){\small \sc $\exists$ algorithm for Sparse polynomial identification};
\end{tikzpicture}
\end{center} 
Such an algorithm $\mathcal A_\text{id}$ is used hereafter that is proposed in (\cite{alamirbook2025}, chapter 10) although the well known \texttt{lassoLars} of scikit-learn \citep{scikit-learn} could have been used as well as far as the number of candidate monomials defined by \eqref{expressionofnc} remains moderate.
\e 
The last definitions we need concerns the metrics of error profiles which are used hereafter in comparing different solutions leading to different error's profiles. More precisely, an error profile $\mathbf e\in \mathbb R^N_+$ representing the profile of reconstruction error of some time-series $\mathbf v$, can be summarized by the following normalized percentiles:
\begin{equation}
p_q\bigl(\mathbf e:=\vert \mathbf v-\hat{\mathbf v}\vert )\bigr):=\dfrac{\texttt{percentile}(\vert \mathbf e\vert, q)}{\epsilon+\texttt{median}(\vert \mathbf v\vert) }\label{defdepqe}
\end{equation}
Given a set of $r$ solutions and their associated error's profiles $\{\mathbf e^{(\ell)}\}_{\ell=1}^r$, denoting by $p^{(\ell)}_q$ the associated percentiles defined by \eqref{defdepqe}, the selected solution is defined by the formulae: 
\begin{equation}
\ell_\star := \texttt{arg}\min_{\ell}\Bigl\{p_{95}^{(\ell)}\quad \vert \quad p_{100}^{(\ell)}\le \eta \times \bigl[\min_{\sigma}p^{(\sigma)}_{100}\bigr]\Bigr\} \label{defdebest}
\end{equation}
where $\eta>1$ is a predefined parameter. For instance, taking $\eta=1.2$ induces a criterion according to which, the \textit{best} solution is the one that shows the minimum $95$-percentiles among all solutions that are not more than $20\%$ worse than the best solution when considering the maximum value of the error.
\section{The methodology}\label{sec-method}
The methodology can be described via the following steps: 
\begin{itemize}
\item[1)] {\sc Generate the datasets}. In this step training, validation and test datasets are built using the presumably known excitation protocol with different sets of excitation parameters (see next section).  \\
\item[2)] {\sc Generate solutions}. Since the degree of the system is not known, one needs to generate solutions for all possible values of $n$ [see \eqref{siso}] ranging from 1 to 4. For each order $\ell$, a solution is generated using the training dataset and the associated error profile $\mathbf e^{(\ell)}$ is computed on the validation set. \\
\item[3)] {\sc Choose the best solution}. Using \eqref{defdebest}, the best derivation order $n=\ell_\star$ is selected and the associated $p_q(\mathbf e^{(n)})$ is computed on the test datasets in order to assess the generalization power of the identified model. \\
\item[4)] {\sc Test feedback design}. In case the results on the test dataset are satisfactory, one disposes of a control-oriented model of the form: 
\begin{equation}
y^{(n)}= \mathcal P(y^{(0)},\dots, y^{(n-1)}, u)\label{defdeynxiu}
\end{equation}
for which a candidate dynamic observer might be given by:
\begin{equation}
\hat y^{(n)}= P(\hat y^{(0)},\dots, \hat y^{(n-1)}, u)+L(y-\hat y^{(0)})\label{obs}
\end{equation}
where the observer gain matrix $L$ is computed for the linear system represented by the chain of $n$ integrators with known leading term. Using the so estimated quantities, any control design is eligible to yield an dynamic output feedback of the form:
\begin{equation}
u = K(\hat y^{(0)}, \dots, \hat y^{(n-1)}, y_\text{ref}) \label{feedback}
\end{equation}
\end{itemize}
Notice that the present paper is mainly concerned with the first three steps leading to the identification of the continuous-time model of the dynamic system. In particular, constrained nonlinear model predictive control can be designed for the control task while a  moving horizon state estimator, high gain or sliding-mode nonlinear state estimators can be implemented. Nevertheless and for the sake of completeness, a concrete instantiation is provided for the illustrative example discussed in the next section.
\section{Experiments}\label{sec-illustration}
\subsection{The unknown system to be identified}
In order to validate the framework proposed in the previous section, let us consider the automotive Electronic Throttle Control (\textbf{ETC}) system given by \citep{CONATSER200423}: 
\begin{subequations}\label{etcsystem}
\begin{align}
\dot x_1&=x_2\label{systa}\\
\dot x_2&=\dfrac{1}{N_m^2J_m+J_g}\Bigl[\phi(x,p)+N_mK_tx_3\Bigr]\label{systb}\\
\dot x_3&=\dfrac{1}{L_a}\Bigl[-N_mK_bx_2-R_ax_3+K_g u\Bigr]\label{systc}
\end{align}
\end{subequations}
where $x:=(\theta, \dot\theta, e_a)$, in which $\theta$ stands for the angle of the admission device while $e_a$ is the electromotor torque  induced by the current $u=i_a$ serving as the control input. The vector $p=(N_m, J_m, J_g,\dots)$ gathers all the parameters involved in \eqref{etcsystem} (see Table \ref{tab_param_etc} for the values). 
\e 
The nonlinear map $\phi(x,p)$ appearing in \eqref{systb} is given by:
\begin{align}
\phi(x,p):=&-K_{sp}(x_1-\pi/2)-(N_m^2b_m+b_t)x_2 - \nonumber \\ &-2P_\text{atm}(\pi-x_1)R_p^2R_{af}\cos^2(x_1),
\end{align}
\begin{table}
\begin{center}
\begin{tabular}{lcccc} \toprule
    {parameter} & {value} & {parameter} & {value} \\ \midrule
    $p_1=N_m$  & 4 &       $p_7=K_t$ &      0.1045 \\
    $p_2=J_m$  & 0.0004 &       $p_8=R_p$  &     0.0015\\
    $p_3=J_g$  & 0.005 &        $p_9=R_{af}$ &         0.002\\
    $p_4=b_m$  & 0.03 &       $p_{10}=L_a$ &       0.003\\
    $p_5=b_t$  & $3.4\times 10^{-3}$ &       $p_{11}=K_b$ &  0.1051\\
    $p_6=K_{sp}$  & $0.4316$ &       $p_{12}=R_a$ &       1.9\\
    $K_g$ & 100 & \ & \\
 \bottomrule
\end{tabular}
\vskip 1mm
\end{center}
\caption{ETC-system's parameters values.} \label{tab_param_etc} 
\end{table}
The measured quantities are the angular position $y=\theta$ and the control input $u$. 
\subsection{The safe excitation input profiles}
\begin{figure}
    \centering
    \includegraphics[width=\linewidth]{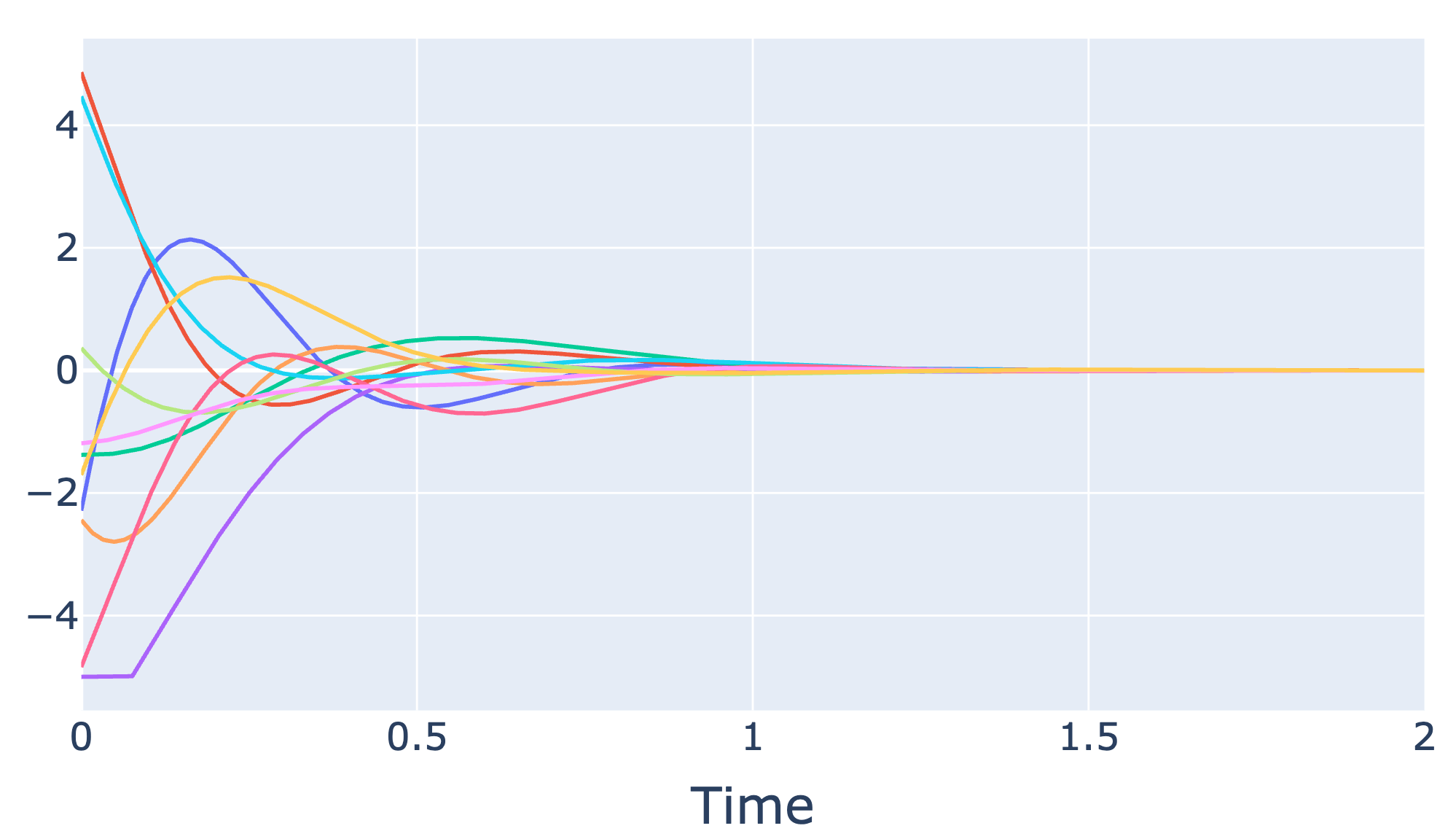}
    \caption{Typical randomly drawn $u$-excitation profiles according to \eqref{defdeut}.}
    \label{fig:excitation}
\end{figure}

The following parameterized set of damped excitation signals is considered for the generation of the datasets:
\begin{equation}
u(t) = \texttt{Sat}_{-\bar u}^{+\bar u}\Bigl(e^{-\mu t}\times \sum_{i=1}^{n_B}\alpha_i\sin\bigl(\omega_it+\psi_i\bigr)\Bigr)\label{defdeut}
\end{equation}
in which $(\alpha_i, \psi_i)$ are randomly generated in $[-\bar u,+\bar u]\times [0,2\pi]$. The pulsation $\omega_i=i\times \omega_\text{max}/n_B$. The values $\omega_\text{max}=10$, $\mu=4$, $\bar u=5$ and $n_B=20$ are used. Typical randomly generated profiles are shown in Fig. \ref{fig:excitation}.
\e 
\subsection{Building working datasets}
A set of 100 damped excitation scenarios such as the ones depicted in Fig. \ref{fig:excitation} have been generated and the resulting measurement profiles collected to constitute the working dataset. Only 25 of these scenarios have been used in the training step while the remaining 75 have been used in the test. With a sampling period of $\tau=0.003$, the complete dataset involved 100,000 rows (samples) among which,  25,000 samples are used in training while 75,000 are used to check the quality of the model on unseen data. 
\e 
A noise-to-signal ratio of $3\%$ has been applied to the measurement profiles leading to the typical noisy measurement profiles shown in Fig. \ref{fig:typical_noise}. More precisely, the measurement $y_m$ fed to the algorithm is obtained from the original noise-free measurement $y$ using:
\begin{equation}
 y_m := y + \bar\eta \times \texttt{percentile}(abs(y), 99)\times \nu \label{noisy}
\end{equation}
where $\bar\nu=0.03$ while $\nu$ is a normalized unitary white noise. 
\subsection{Fitting polunomial models}
\begin{figure}
    \centering
    \includegraphics[width=\linewidth]{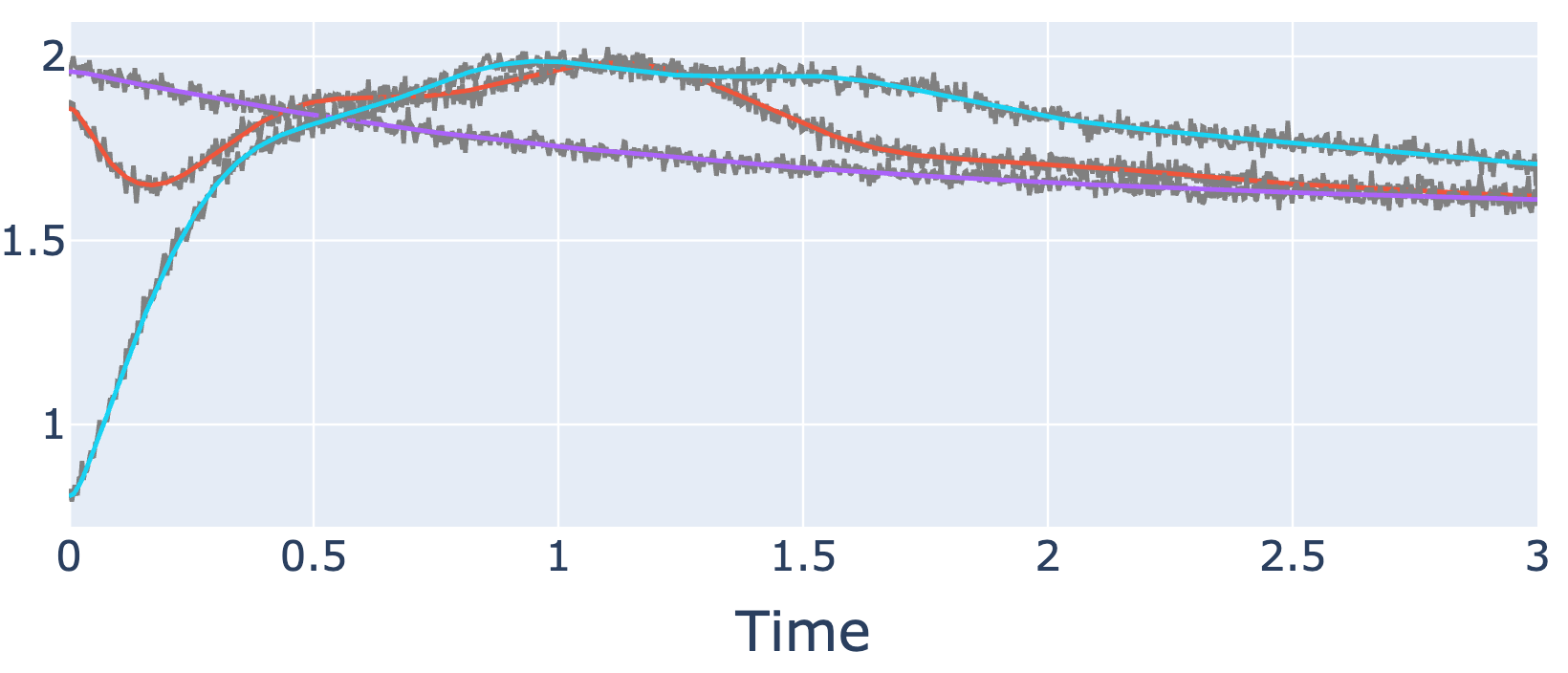}
    \caption{Typical noisy measurements used in the training dataset to fit the sparse polynomial models.}
    \label{fig:typical_noise}
\end{figure}

Tables \ref{dfe_train_0} and \ref{dfe_train_03} show the fitting errors (defined by \eqref{defdepqe} in which $\epsilon=10^{-12}$ is used) respectively when using noise-free or noisy measurements. Each table shows the error statistics for different model's order involved in \eqref{siso}. These statistics clearly show that the third order model $n=3$ is the only appropriate ones, which is expected from our knowledge of the hidden ODEs generating the data. 
\e The results also show the rather nice robustness of the approach to the presence of decently high level of noise (see Fig. \ref{fig:typical_noise}) except for a significant increase of the maximum absolute error which nevertheless remains lower that 80\% of the median value of the third derivative's amplitude. This is obviously a sort of incompressible instant-wise error due to the presence of measurement and does not necessarily mean that the model is erroneous as the latter should capture only the raw signal and not the noisy one. 

\begin{table}
\begin{center}
\begin{tabular}{lrrrr}
\toprule
$q$ & n=1 & n=2 & n=3 & n=4 \\
\midrule
50\% & 0.36 & 1.12 & \textbf{0.01} & 0.25 \\
80\% & 0.92 & 1.77 & \textbf{0.01} & 0.56 \\
90\% & 1.75 & 2.81 & \textbf{0.03} & 0.88 \\
95\% & 3.57 & 4.82 & \textbf{0.05} & 1.54 \\
98\% & 7.31 & 8.03 & \textbf{0.08} & 2.63 \\
99\% & 9.28 & 10.96 & \textbf{0.10} & 3.30 \\
100\% & 18.69 & 26.92 & \textbf{0.17} & 7.76 \\
\bottomrule
\end{tabular}

\end{center}
\vskip 1mm 
\caption{\textbf{Noise-free measurements}. Normalized percentiles of the absolute identification errors on the train data for different candidate model's orders $n\in \{1,2,3,4\}$. $d=3$.}\label{dfe_train_0}
\end{table}

\begin{table}
\begin{center}
\begin{tabular}{lrrrr}
\toprule
$q$ & n=1 & n=2 & n=3 & n=4 \\
\midrule
50\% & 0.45 & 1.32 & \textbf{0.01} & 0.49 \\
80\% & 1.11 & 2.01 & \textbf{0.04} & 1.04 \\
90\% & 1.89 & 3.26 & \textbf{0.06} & 1.79 \\
95\% & 3.58 & 5.52 & \textbf{0.08} & 2.75 \\
98\% & 7.34 & 8.99 & \textbf{0.11} & 4.78 \\
99\% & 9.21 & 11.78 & \textbf{0.12} & 6.16 \\
100\% & 18.41 & 28.33 & \textbf{0.82} & 37.01 \\
\bottomrule
\end{tabular}

\end{center}
\vskip 1mm 
\caption{\textbf{Noisy measurements}. Normalized percentiles of the absolute identification errors on the train data for different candidate model's orders $n\in \{1,2,3,4\}$, $d=3$.}\label{dfe_train_03}
\end{table}

\e Now selecting the best model (the one with order $n=3$), one can check its generalization power by examining the same statistics of error on the unseen samples included in the test dataset. The results are shown in Table \ref{dfetest} which demonstrates a good generalization power as the order of magnitude of the remains practically the same up to 99\% of the data while a noticeable increase in the maximum value. 
\e The same results are shown in Table \ref{dfetest_d1} in the case where a polynomial of degree 1 is used to fit the data. Despite of a noticeably higher error statistics compared to the ones obtained with a polynomial of degree 3 (Table \ref{dfetest}), the error's statistics remain quite small and makes this model eligible to be used in the design of the feedback control. Notice however that even if the linear polynomial is used to design the control law, this still induces a nonlinear control as the second derivatives $y^{(2)}=\dot x_2$, given by the r.h.s of \eqref{systb}, is obviously a nonlinear function of the state. 

\begin{table}
\begin{center}
\begin{tabular}{lrrrrrrr}
\toprule
$q$ & 50\% & 80\% & 90\% & 95\% & 98\% & 99\% & 100\% \\
\midrule
Error & \textbf{0.01} & \textbf{0.03} & \textbf{0.04} & \textbf{0.06} & \textbf{0.12} & \textbf{0.22} & \textbf{1.46} \\
\bottomrule
\end{tabular}

\end{center}
\vskip 1mm 
\caption{\textbf{Noisy measurements}. Normalized percentiles of errors on \textbf{test dataset} using the model's order $n=3$, $d=3$.}\label{dfetest} 
\end{table}

\begin{table}
\begin{center}
\begin{tabular}{lrrrrrrr}
\toprule
q & 50\% & 80\% & 90\% & 95\% & 98\% & 99\% & 100\% \\
\midrule
Error & 0.03 & 0.07 & 0.09 & 0.13 & 0.24 & 0.36 & 1.82 \\
\bottomrule
\end{tabular}
\end{center}
\vskip 1mm 
\caption{\textbf{Noisy measurements}. Normalized percentiles of errors on \textbf{test dataset} using the model's order $n=3$, $d=1$.}\label{dfetest_d1}
\end{table}

\subsection{Feedback design}
\begin{figure}
    \centering
    \includegraphics[width=\linewidth]{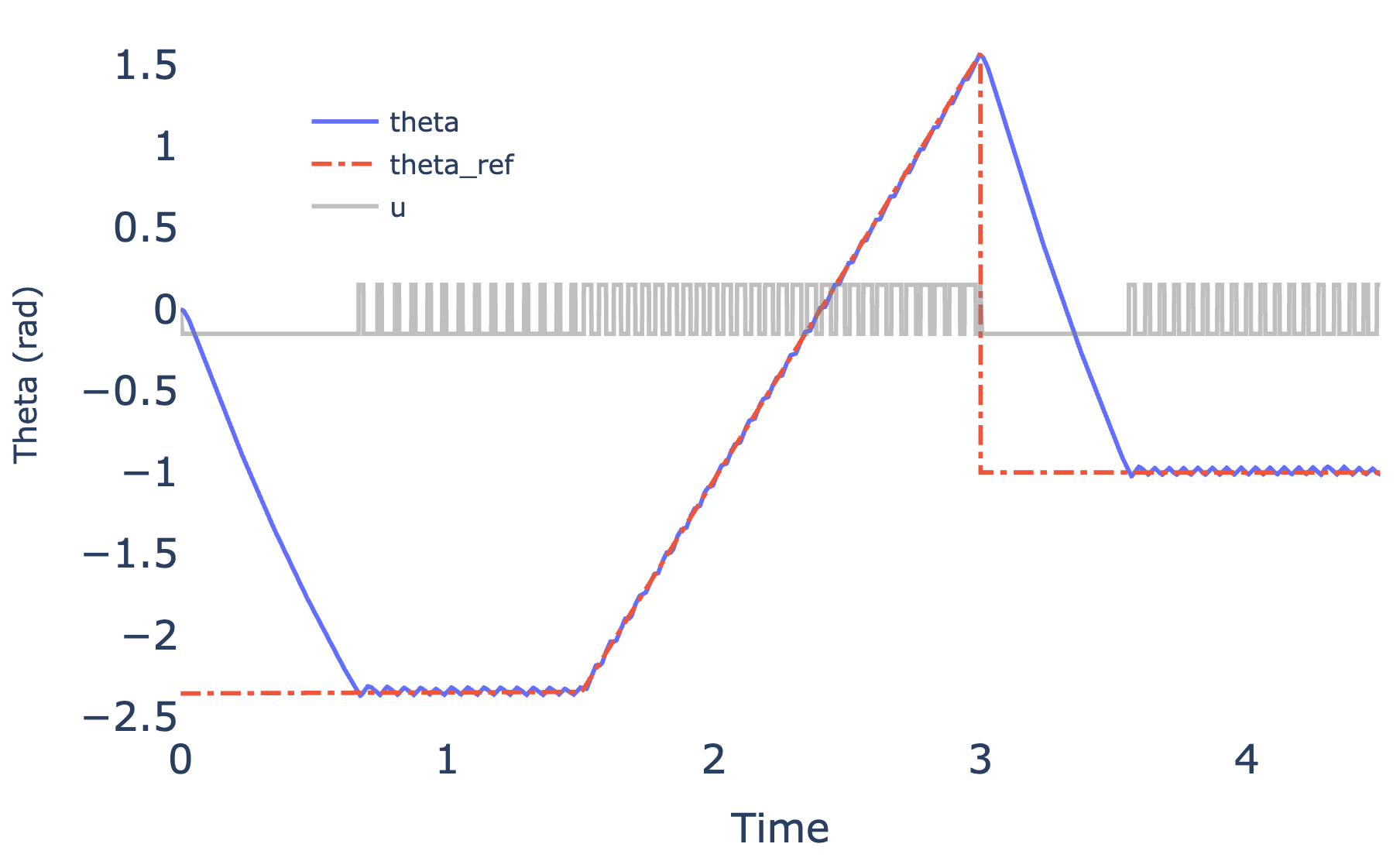}
    \caption{Behavior of the closed-loop control using a sampling period of $\tau=800\ \mu sec$ and a blocking order $\kappa=3$.}
    \label{fig:closed-loop}
\end{figure}

The control design is based on a very specific version of NMPC that is tailored to this specific problem of regulating a chain of 3 integrators. This can be summarized by the following two steps: \e 
1) At each decision instant $k$, compute a trajectory that is compatible with the estimated values $\hat y^{(i)}(k)$, $i=0,\dots,2$ and the targeted desired value $y_\text{ref}$. This is done by identifying the coefficients $\alpha_i$ of the following time-profile:
\begin{equation}
\bar y = \Psi(t, \alpha):=\sum_{j=1}^{4} \alpha_i\times e^{-\sigma j t} \label{defdePsi}
\end{equation}
through the solution of the following linear system of equations in the unknown $\alpha$:
\begin{equation}
\begin{bmatrix} 
\Psi(0,\alpha)\cr 
\dot \Psi(0,\alpha)\cr 
\ddot \Psi(0,\alpha)\cr 
\Psi(\infty,\alpha)
\end{bmatrix} = \begin{bmatrix} 
\hat y^{(0)}_k \cr 
\hat y^{(1)}_k \cr 
\hat y^{(2)}_k \cr 
y_\text{ref}
\end{bmatrix} =:\begin{bmatrix} 
\hat \xi_k\cr 
y_\text{ref}
\end{bmatrix}
 \label{defdexihat}
\end{equation}
where $\xi$ is used to denote the state vector of the system in the derivatives-based coordinates while $\hat\xi_k$ denotes the observer based estimation of $\xi_k$. The solution of this system leads to the $\hat\xi_k$-dependent solution denoted by $\alpha(\hat\xi_k)$.
\e
2) The so obtained solution $\alpha(\hat\xi_k)$ when used in \eqref{defdePsi} enables to compute a \textit{desired} third derivative $\Psi^{(3)}\Bigl(0,\alpha(\hat\xi_k)\Bigr)$ that one can try to achieve using the appropriate control $u$ and the identified polynomial given by \eqref{defdeynxiu}, namely:
\begin{equation}
u^\star(\hat\xi_k) = \text{arg}\min_{v\in \mathbb U} \left\vert \Psi^{(3)}\Bigl(0,\alpha(\hat\xi_k)\Bigr) - \mathcal P(\hat\xi_k,v)\right\vert\label{}
\end{equation}
In order to get a real-time implementable control for this sensitive system (that needs very small sampling periods), this process is not necessarily performed at each sampling period (set here to $\tau=800\mu sec$) as this might lead to a computation time exceeding $\tau$. Rather, we block the control over an integer number $\kappa$ of sampling periods. Moreover, the polynomial of order 1 (Table \eqref{dfetest_d1}) is used for the sake of reducing the computation time and getting simpler design. Closed-loop simulation is shown in Fig. \ref{fig:closed-loop} for a 5 sec scenario involving step and ramp like changes in the reference angle with a blocking parameter of $\kappa=3$. The control saturation level is set to $\bar u=0.15$ (using higher actions lead to larger chattering around the reference). This induces a real-life compatible feedback (the total simulation time of the 5 sec scenario is 1.3 sec. 

\section{Conclusion and future work}\label{sec-conc}
In this paper, a framework for the identification of continuous-time nonlinear system is proposed that leverages some recent advances in the high derivative computation and multivariate polynomial identification. Undergoing investigation concern the application of this framework to a set of benchmark including MIMO systems. Moreover, a freely available python module is under preparation for potential interested users. 



\bibliography{ifacconf}             
\end{document}